\documentclass[conference,9pt]{IEEEtran}
\IEEEoverridecommandlockouts

\usepackage{
amsmath,
graphicx,
cite,
multirow,
booktabs,
amssymb,
enumitem,
url,
}

\begin{document}

\title{Channel-Aware Domain-Adaptive Generative Adversarial Network for Robust Speech Recognition}

\author{
\IEEEauthorblockN{
Chien-Chun Wang\IEEEauthorrefmark{1},
Li-Wei Chen\IEEEauthorrefmark{3},
Cheng-Kang Chou\IEEEauthorrefmark{3},
Hung-Shin Lee\IEEEauthorrefmark{3},
Berlin Chen\IEEEauthorrefmark{1}, and
Hsin-Min Wang\IEEEauthorrefmark{2}
}
\IEEEauthorblockA{
\IEEEauthorrefmark{1} 
Dept. Computer Science and Information Engineering, National Taiwan Normal University, Taiwan
}
\IEEEauthorblockA{
\IEEEauthorrefmark{2}
Institute of Computer Science, Academia Sinica, Taiwan
}
\IEEEauthorblockA{
\IEEEauthorrefmark{3}
United-Link Co., Ltd., Taiwan
}
}

\maketitle

\begin{abstract}

While pre-trained automatic speech recognition (ASR) systems demonstrate impressive performance on matched domains, their performance often degrades when confronted with channel mismatch stemming from unseen recording environments and conditions. To mitigate this issue, we propose a novel channel-aware data simulation method for robust ASR training. Our method harnesses the synergistic power of channel-extractive techniques and generative adversarial networks (GANs). We first train a channel encoder capable of extracting embeddings from arbitrary audio. On top of this, channel embeddings are extracted using a minimal amount of target-domain data and used to guide a GAN-based speech synthesizer. This synthesizer generates speech that faithfully preserves the phonetic content of the input while mimicking the channel characteristics of the target domain. We evaluate our method on the challenging Hakka Across Taiwan (HAT) and Taiwanese Across Taiwan (TAT) corpora, achieving relative character error rate (CER) reductions of 20.02\% and 9.64\%, respectively, compared to the baselines. These results highlight the efficacy of our channel-aware data simulation method for bridging the gap between source- and target-domain acoustics.

\end{abstract}


\section{Introduction}

Automatic speech recognition (ASR) has become an indispensable technology, powering applications ranging from virtual assistants to transcription services. Recent advancements in deep learning, particularly with architectures such as convolutional neural networks (CNNs) \cite{pan2020}, long short-term memory networks (LSTMs) \cite{park2020,zeyer2021}, Transformers \cite{baevski2020,haidar2021,xu2021,radford2022}, and Conformers \cite{gulati2020,chan2021,kim2022}, have significantly improved ASR accuracy across various conditions. However, these architectures often remain susceptible to performance degradation caused by channel mismatch---a discrepancy in acoustic characteristics between training and test data due to differences in recording equipment.

This vulnerability is particularly evident in scenarios like teleconferencing, where diverse microphones, ranging from professional condenser microphones to built-in webcams, introduce significant variations in signal quality. This mismatch can drastically degrade performance. For instance, as shown in Table \ref{tab:channels_eval}, using Whisper\textsubscript{Tiny} \cite{radford2022} fine-tuned on Condenser data from the Hakka Across Taiwan (HAT) \cite{liao2023} and Taiwanese Across Taiwan (TAT) \cite{liao2022} corpora results in drastically increased character error rates (CERs) when evaluated on other microphone types, highlighting the urgent need for more channel-robust ASR systems.

To address this challenge, researchers have explored domain adaptation techniques \cite{hsu2018,mun2019,li2020,wang2022,wang2023,yang2024} that aim to bridge the gap between training and test distributions. While these techniques have shown promise, they often involve complex training procedures or may not fully exploit the underlying relationship between domains. Recently, data simulation has emerged as an alternative approach \cite{hu2018,chen2022,chen2023}, generating synthetic target-domain data from source-domain data to facilitate model adaptation without requiring paired samples. However, existing data simulation approaches primarily focus on mitigating noise-related mismatches and lack the sophistication to effectively tackle channel discrepancies. This highlights a crucial need for novel methods specifically designed to enhance ASR robustness across diverse recording channels.

\begin{table}[t]
\small
\caption{CERs (\%) with respect to various test channels.}
\vspace{-5pt}
\label{tab:channels_eval}
\centering
\setlength{\tabcolsep}{20pt}
\begin{tabular}{lcc}
\toprule
\bf Test Channels & \bf HAT & \bf TAT \\
\toprule
\bf Condenser & \bf 2.43 & \bf 9.39 \\
Lavalier & 2.94 & 9.56 \\
iPhone & 4.10 & 11.21 \\
Android & 4.70 & 12.76 \\ 
\bottomrule
\end{tabular}
\vspace{-15pt}
\end{table}

\begin{figure*}[ht]
\centering
\includegraphics[width=0.95\linewidth]{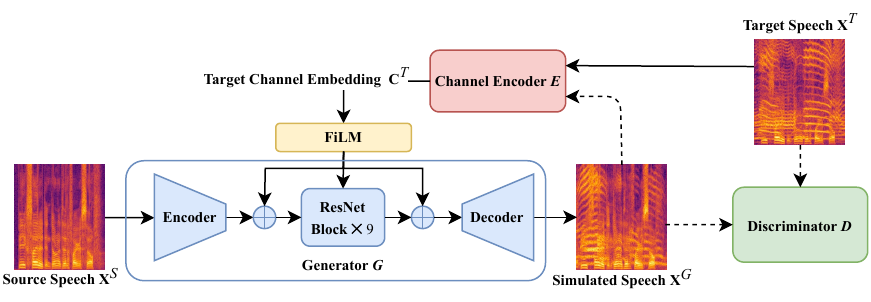}
\vspace{-5pt}
\caption{The architecture of our proposed method, CADA-GAN. The dotted arrows indicate that during the training phase, simulated speech $\mathbf{X}^G$ is used together with target speech $\mathbf{X}^T$ to 1) train the discriminator, and 2) contribute to channel reconstruction. The $\bigoplus$ operator denotes element-wise tensor addition.}
\label{fig:CADA-GAN}
\vspace{-10pt}
\end{figure*}

To overcome the limitations of existing approaches, we introduce CADA-GAN, a novel Channel-Aware Domain-Adaptive Generative Adversarial Network, designed to enhance ASR robustness to channel mismatch. Our method leverages a two-step process: channel embedding extraction and domain-adaptive speech synthesis.

First, a channel encoder is trained to extract detailed channel embeddings from target-domain speech. These embeddings capture the unique acoustic characteristics of the target recording environment. Next, a GAN architecture, guided by the extracted embeddings, learns a source-to-target domain transformation. The generator synthesizes speech that accurately replicates target-domain channel characteristics while preserving the phonetic content of the source speech. The discriminator ensures the authenticity of the generated samples, further refining the domain adaptation process.

Moreover, CADA-GAN requires only a small amount of unpaired target-domain data during training, making it highly practical for real-world scenarios. At inference time, only the generator and channel encoder are utilized, eliminating the need for additional transcriptions. The generator synthesizes abundant training data by randomly pairing source speech with the limited target speech, inheriting the source transcriptions. This data augmentation strategy allows for fine-tuning any ASR model to significantly improve channel robustness.

Unlike previous GAN-based approaches, such as UNA-GAN \cite{chen2023}, which adopt a global approach to domain adaptation, CADA-GAN generates customized, domain-specific speech by explicitly conditioning the generation process on channel embeddings extracted from individual target utterances. This fine-grained control enables it to precisely align the synthesized speech with the target channel characteristics. We evaluate CADA-GAN on the challenging HAT and TAT corpora. Our method demonstrates its effectiveness by achieving relative CER reductions of 20.02\% and 9.64\%, respectively, compared to the baseline trained solely on the source domain.

\section{Proposed Method}\label{sec:method}

Fig. \ref{fig:CADA-GAN} illustrates the architecture of our proposed CADA-GAN, which comprises three key components: a generator ($G$), a discriminator ($D$), and a channel encoder ($E$). The process begins with the channel encoder, which extracts a channel embedding ($\mathbf{C}^T$) from a target-domain spectrogram ($\mathbf{X}^T$). This embedding encapsulates the distinct acoustic characteristics of the target recording environment. The generator then utilizes this embedding alongside a source-domain spectrogram ($\mathbf{X}^S$) to synthesize a simulated spectrogram ($\mathbf{X}^G$) that mimics the target-domain channel characteristics while preserving the phonetic content of the source speech. Finally, the discriminator distinguishes between real target spectrograms and simulated spectrograms, providing feedback to the generator during training.

\subsection{Generator and Discriminator}\label{sec:gan}

The generator ($G$) is designed to transform a source-domain spectrogram ($\mathbf{X}^S$) into a simulated target-domain spectrogram ($\mathbf{X}^G$). It achieves this by first processing the source spectrogram through two 2D downsampling convolutional layers (kernel size: $3\times 3$, stride: $2\times 2$), followed by nine residual blocks to capture deep, hierarchical representations. Each residual block consists of two convolutional layers (kernel size: $3\times 3$, stride: $1\times 1$) and a dropout layer to prevent overfitting. Finally, two transposed convolutional layers (kernel size: $3\times 3$, stride: $2\times 2$) upsample the learned representations to generate the simulated spectrogram.

The discriminator ($D$) plays a crucial role in ensuring the authenticity of the generated spectrograms. It distinguishes between real target spectrograms ($\mathbf{X}^T$) and simulated ones using five 2D convolutional layers (kernel size: $4\times 4$) with Leaky ReLU activation functions. The stride is set to $2\times 2$ for the first three layers and $1\times 1$ for the last two, gradually increasing the receptive field. The adversarial loss employed during training is defined as follows:
\begin{equation}
\begin{split}
\mathcal{L}_{adv}(G,&D,\mathbf{X}^T,\mathbf{X}^S,\mathbf{C}^T) = \mathbb{E}_{\mathbf{x} \sim \mathbf{X}^T} \left[ \log D(\mathbf{x}) \right] \\
&+ \mathbb{E}_{\mathbf{x} \sim \mathbf{X}^S, \mathbf{c} \sim \mathbf{C}^T} \left[ \log (1-D(G(\mathbf{x},\mathbf{c}))) \right].
\end{split}
\end{equation}
This adversarial loss encourages the generator to produce spectrograms that closely resemble real target spectrograms, while the discriminator learns to identify subtle differences that distinguish real from simulated data. This adversarial training process compels the generator to continuously improve its ability to generate realistic and domain-specific speech.

\subsection{Channel Encoder}\label{sec:encoder}

Drawing inspiration from recent advancements in ``aware'' techniques \cite{li2021,wang2023,hu2024}, we introduce a dedicated channel encoder ($E$) to extract channel embeddings ($\mathbf{C}^T$) from the final layer of a pre-trained model. Unlike conventional approaches that directly utilize raw spectrograms as channel information, our channel encoder focuses on capturing high-level, discriminative channel characteristics.

Specifically, our channel encoder leverages a MFA-Conformer model \cite{zhang2022}, which is pre-trained on the HAT corpus \cite{liao2023}. The training data consists of recordings from speakers uttering identical content using different microphones at the same time, ensuring that the model is unaffected by speech content or speaker identity. By excluding both the source and target channels used in the main experiment, we enhance the model's ability to classify various channels based purely on their acoustic properties. This strategy enables the channel encoder to effectively capture detailed channel characteristics without incorporating phonetic information, leading to more robust and generalizable channel embeddings.

The channel embeddings are integrated to the generator using Feature-wise Linear Modulation (FiLM) \cite{perez2018}. The embeddings undergoes separate linear transformations to produce weights and biases. These are used to modulate the output features from specific layers in the generator. To further ensure that the generated spectrograms accurately reflect the target-domain channel characteristics, we introduce a channel reconstruction loss:
\begin{equation}\label{eq_ch}
\mathcal{L}_{ch}(G,\mathbf{X}^S,\mathbf{C}^T) = \mathbb{E}_{\mathbf{x} \sim \mathbf{X}^S, \mathbf{c} \sim \mathbf{C}^T} \left[ \left\| \mathbf{c} - E(G(\mathbf{x},\mathbf{c})) \right\|_1 \right].
\end{equation}
This loss function encourages the generator to synthesize spectrograms that, when processed by the channel encoder, yield embeddings highly similar to the original target channel embeddings. This reinforces the channel awareness of the generator, ensuring that the generated speech accurately captures the subtle nuances of the target recording environment.

\subsection{Patch-wise Contrastive Learning}\label{sec:pcl}

To maintain linguistic consistency between the simulated and source speech, we apply patch-wise contrastive learning \cite{park2020a}. This approach maximizes mutual information, particularly shared speech content, between source and simulated spectrograms. Specifically, we utilize the generator to extract deep features from both spectrograms. A small patch from the simulated representation serves as the ``query'', with the corresponding patch from the source as the ``positive'' sample. We select 256 patches from the source as ``negative'' samples. These patches are projected into a lower-dimensional space using two linear layers with 256 units each and ReLU activation. The contrastive loss, computed across five generator layers, measures the cross-entropy loss between the ``query'' patch and both positive and negative patches. This encourages high similarity between corresponding patches in the source and simulated spectrograms, while distinguishing them from random patches. The loss is defined as:
\begin{equation}
\resizebox{0.5\textwidth}{!}{
$\mathcal{L}_{pcl}(G,\mathbf{X}^S) = \sum\limits_{l=1}^{L} \sum\limits_{i=1}^{I} -\log \frac{e^{\left( \hat{z}_l^i \cdot z_l^i / \tau \right)}}{e^{\left( \hat{z}_l^i \cdot z_l^i / \tau \right)} + \sum_{j=1}^{J} e^{\left( \hat{z}_l^i \cdot z_l^j / \tau \right)}},$
}
\end{equation}
where $z_{l}^{i}$ represents the $i^{th}$ positive patch from source representations at the $l^{th}$ layer of the generator, $\hat{z}_{l}^{i}$ denotes the corresponding patch from simulated representations, and $z_{l}^{j}$ refers to the $j^{th}$ negative patch from simulated representations at the same layer. The temperature parameter $\tau$ regulates the contrastive learning process. This loss function is applied to both source ($\mathcal{L}_{pcl}(G,\mathbf{X}^S)$) and target ($\mathcal{L}_{pcl}(G,\mathbf{X}^T)$) spectrograms to maintain consistent speech content and minimize unnecessary changes.

\subsection{Training Objective and Adaptation Process}\label{sec:train & adapt}

The training objective is to optimize GAN using a comprehensive loss function that includes the adversarial loss, patch-wise contrastive learning losses for both source and target spectrograms, and the channel reconstruction loss. The total loss function is defined as:
\begin{equation}\label{eq_total}
\begin{split}
\mathcal{L}_{total} = &\mathcal{L}_{adv}(G,D,\mathbf{X}^T,\mathbf{X}^S,\mathbf{C}^T) + \mathcal{L}_{pcl}(G,\mathbf{X}^S) \\
&+ \mathcal{L}_{pcl}(G,\mathbf{X}^T) + \lambda_{ch}\mathcal{L}_{ch}(G,\mathbf{X}^S,\mathbf{C}^T),
\end{split}
\end{equation}
where $\lambda_{ch}$ weights the channel reconstruction loss.

To address limited target-domain data, an equal amount of speech is randomly sampled from the source domain. The model is trained on this unpaired dataset, optimizing with the total loss function $\mathcal{L}_{total}$ as specified in (\ref{eq_total}). After training, the generator serves as a domain converter $F^{S \sim T}$, transforming $\mathbf{X}^S$ to $\mathbf{X}^T$, with the pre-trained channel encoder aiding in channel simulation. This simulation uses plentiful source speech and randomly selected target speech from training. The augmented data enhances the fine-tuning of ASR models without requiring additional transcriptions.

\section{Experiments}\label{sec:exp}

We conducted extensive experiments on two benchmark datasets to evaluate the efficacy of our proposed CADA-GAN method for domain-adaptive ASR.

\textbf{HAT} \cite{liao2023}: The HAT corpus comprises hundred thousands sets of recordings, where each set was uttered by the same speaker with identical speech content using eight different microphones, reflecting diverse recording conditions. The recorders include an \textbf{iPhone}, an \textbf{Android} phone, a \textbf{Webcam}, a professional \textbf{Condenser} microphone, a \textbf{Lavalier} microphone, a cheap PC microphone (\textbf{PC-Mic}), and an X-Y stereo microphone (\textbf{ZOOM-X} and \textbf{ZOOM-Y}). The dataset contains 97,385 training sets (779,080 utterances in total) and 4,559 test sets (36,472 utterances in total). We selected recordings from \textbf{Condenser} as the source domain and those from \textbf{Webcam} as the target domain due to their significant acoustic mismatch. To train our GAN, we randomly sampled 40 utterances from each domain, demonstrating the method's effectiveness with limited target-domain data.

\textbf{TAT} \cite{liao2022}: To further validate that our channel encoder does not inadvertently capture phonetic information, we conducted additional experiments on the TAT corpus. This dataset is similar to HAT but excludes recordings from \textbf{Webcam} and \textbf{PC-Mic}. We used \textbf{Condenser} recordings as the source domain and \textbf{Android} recordings as the target domain for this evaluation. There is no information indicating that HAT and TAT use the same type and brand of devices.

\begin{table}[t]
\small
\caption{CERs (\%) and their relative CER reductions (Rel. \%) on HAT and TAT with respect to various methods.}
\vspace{-5pt}
\label{tab:cer}
\centering
\setlength{\tabcolsep}{10pt}
\begin{tabular}{lcccc}
\toprule
\multirow{2}{*}{\textbf{Model}} & \multicolumn{2}{c}{\textbf{HAT}} & \multicolumn{2}{c}{\textbf{TAT}} \\
\cmidrule(lr){2-3}
\cmidrule(lr){4-5}
 & \bf{CER} & \bf{Rel.} & \bf{CER} & \bf{Rel.} \\ 
\toprule
Vanilla ASR & 10.24 & - & 12.76 & - \\ 
UNA-GAN \cite{chen2023} & 9.76 & 4.69 & 11.82 & 7.37 \\ 
\textbf{CADA-GAN} & \textbf{8.19} & \textbf{20.02} & \textbf{11.53} & \textbf{9.64} \\
\midrule
Topline ASR & 3.88 & 62.11 & 10.30 & 19.28\\
\bottomrule
\end{tabular}
\vspace{-15pt}
\end{table}

\subsection{Backbone ASR Models}\label{sec:whisper}

We employ Whisper \cite{radford2022}, an ASR system developed by OpenAI, as our downstream evaluation model. Whisper, based on a large-scale Transformer architecture, is trained on a massive dataset of 680,000 hours of multilingual speech data, showcasing robust performance across various languages and acoustic conditions. Given the resource constraints of edge devices, we specifically chose Whisper\textsubscript{Tiny} as our downstream ASR model. This lightweight version maintains the impressive performance of Whisper while being optimized for deployment on devices with limited computational resources.

To comprehensively evaluate the impact of our proposed method, we trained four variants of the ASR model:
\begin{enumerate}[noitemsep,leftmargin=*]
\item \textbf{Vanilla ASR (Baseline)}: The Whisper\textsubscript{Tiny} fine-tuned solely on the source-domain data without any channel compensation.
\item \textbf{UNA-GAN}: The Whisper\textsubscript{Tiny} fine-tuned using a dataset generated by UNA-GAN \cite{chen2023}, representing a recent baseline approach.
\item \textbf{CADA-GAN}: The Whisper\textsubscript{Tiny} fine-tuned with the augmented dataset generated by CADA-GAN, incorporating both source and simulated target-domain data.
\item \textbf{Topline ASR}: The Whisper\textsubscript{Tiny} fine-tuned directly on the target-domain data, simulating an ideal scenario with abundant labeled target-domain data.
\end{enumerate}

\subsection{Configuration}\label{sec:config}

To ensure effective channel information extraction, we segmented the input speech spectrograms into frames of $129\times 128$ dimensions. We trained the CADA-GAN model for 400 epochs, optimizing the balance between adversarial training and channel reconstruction with a weighting factor of $\lambda_{ch}=0.5$. The Adam optimizer \cite{kingma2015} was employed, with an initial learning rate of 0.0002, for stable training. For domain adaptation, the Whisper\textsubscript{Tiny} model was fine-tuned for 10 epochs using a learning rate of 0.0001.

\section{Results and Discussion}\label{sec:results}

We present the results of our experiments, comparing the performance of CADA-GAN with the baseline models on both HAT and TAT corpora. Additionally, ablation studies were conducted to analyze the contribution of individual components in our method.

\subsection{Main Results on HAT and TAT}\label{sec:cer}

Table \ref{tab:cer} presents the CERs achieved by all ASR models on the HAT and TAT corpora. CADA-GAN demonstrates substantial improvements over the baseline approaches, achieving a remarkable 20.02\% relative CER reduction on the HAT corpus and a 9.64\% relative CER reduction on the TAT corpus compared to the Vanilla ASR model. These results highlight the effectiveness of incorporating the pre-trained channel encoder within the domain adaptation framework.

Furthermore, the consistent performance gains observed on both HAT and TAT, despite the channel encoder being trained solely on the HAT corpus, underscores its ability to capture and leverage channel-specific features effectively while remaining agnostic to phonetic information. This crucial capability enables the model to generalize well across diverse languages and unseen channel conditions.

\begin{table}[t]
\small
\caption{CERs (\%) on HAT and TAT with respect to ablation studies.}
\vspace{-5pt}
\label{tab:ablation}
\centering
\setlength{\tabcolsep}{23pt}
\begin{tabular}{lcc}
\toprule 
\textbf{Model} & \textbf{HAT} & \textbf{TAT} \\
\toprule
\textbf{CADA-GAN} & \textbf{8.19} & \textbf{11.24} \\
- $\mathcal{L}_{ch}$ & 8.77 & 11.59 \\
- Embeddings & 9.05 & 11.65 \\
\bottomrule
\end{tabular}
\vspace{-5pt}
\end{table}

\begin{figure}[t]
\centering
\includegraphics[width=0.95\linewidth]{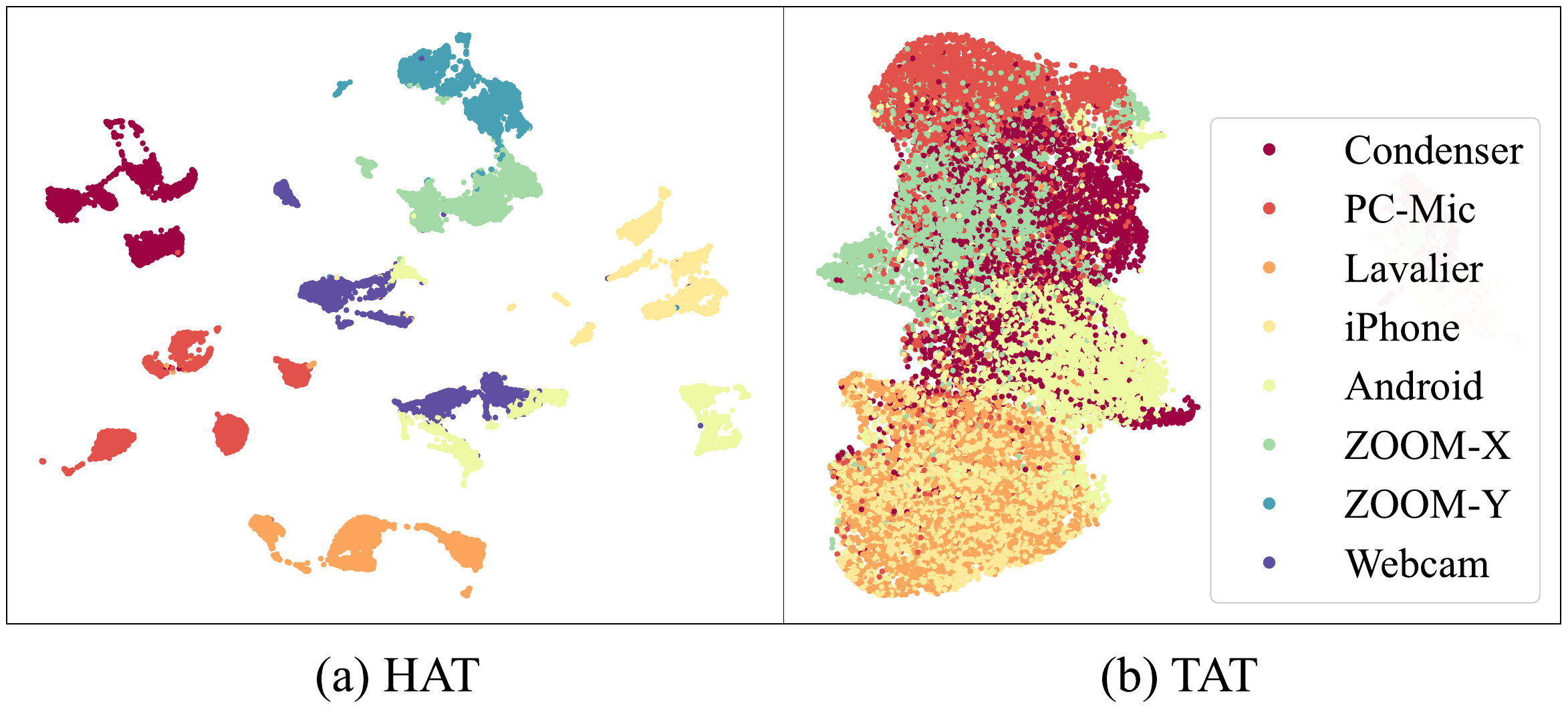}
\vspace{-5pt}
\caption{The UMAP visualization of channel embeddings extracted from eight channel types in the HAT corpus and six channel types in the TAT corpus.}
\label{fig:umap}
\vspace{-5pt}
\end{figure}

\subsection{Ablation Studies}\label{sec:ablation}

To delve deeper into the contribution of each component within CADA-GAN, we conducted ablation studies, summarized in Table \ref{tab:ablation}. Removing the channel reconstruction loss (- $\mathcal{L}_{ch}$) during training resulted in a marginal performance decline, suggesting that while this loss aids in maintaining channel fidelity, its overall impact on speech recognition accuracy is relatively small.

Conversely, omitting the channel embeddings (- Embeddings) during the generation process led to a significant drop in performance. This substantial decrease emphasizes the critical role channel embeddings play in accurately capturing and transferring channel-specific characteristics from the target domain, even though our channel encoder was trained solely on the HAT corpus. These embeddings are essential for enhancing the model's robustness and enabling accurate speech recognition across different channel conditions.

\subsection{UMAP Visualization of Channel Embeddings}\label{sec:umap}

To gain further insights into the workings of CADA-GAN, we visualized the learned channel embeddings and evaluate the perceptual quality of the generated speech. Uniform Manifold Approximation and Projection (UMAP) \cite{mclnnes2018} was employed to visualize the channel embeddings extracted from the HAT and TAT corpora, as shown in Fig. \ref{fig:umap}. In Fig. \ref{fig:umap} (a), a clear separation between different channel types in the HAT corpus is observed. This distinct clustering demonstrates the effectiveness of our pre-trained channel encoder in capturing unique acoustic characteristics associated with each microphone. While the separation is less pronounced in Fig. \ref{fig:umap} (b) for the TAT corpus (where the encoder was not specifically trained), similar channel types are still effectively grouped together. This observation highlights the generalization ability of our channel encoder across different languages, reinforcing its capacity to learn channel-specific features rather than language-dependent patterns.

\subsection{MOS Evaluation on Simulated Data}\label{sec:mos}

To assess the perceptual realism of the generated speech, we conducted a Mean Opinion Score (MOS) evaluation focused specifically on channel characteristics. Ten participants rated the similarity of the perceived recording channel between generated audio samples and target-channel reference recordings on a scale of 1 to 5, with 5 representing the highest similarity. Table \ref{tab:mos} summarizes the MOS scores for each method. CADA-GAN achieves a significantly higher MOS compared to UNA-GAN, indicating that our method generates speech that more closely resembles the target domain in terms of channel characteristics, despite our channel encoder not being explicitly trained on the TAT corpus. Moreover, the smaller standard deviation observed for CADA-GAN suggests a higher consistency in the perceptual quality of the generated speech.

\begin{table}[t]
\small
\caption{MOSs of simulated speech on HAT and TAT.}
\vspace{-5pt}
\label{tab:mos}
\centering
\setlength{\tabcolsep}{13pt}
\begin{tabular}{lcc}
\toprule 
\textbf{Model} & \textbf{HAT} & \textbf{TAT} \\
\toprule
UNA-GAN \cite{chen2023} & 2.90 $\pm$ 0.75 & 2.55 $\pm$ 1.11 \\
\textbf{CADA-GAN} & \textbf{4.06 $\pm$ 0.71} & \textbf{3.09 $\pm$ 1.06} \\
\bottomrule
\end{tabular}
\vspace{-5pt}
\end{table}

\begin{figure}[t]
\centering
\includegraphics[width=0.95\linewidth]{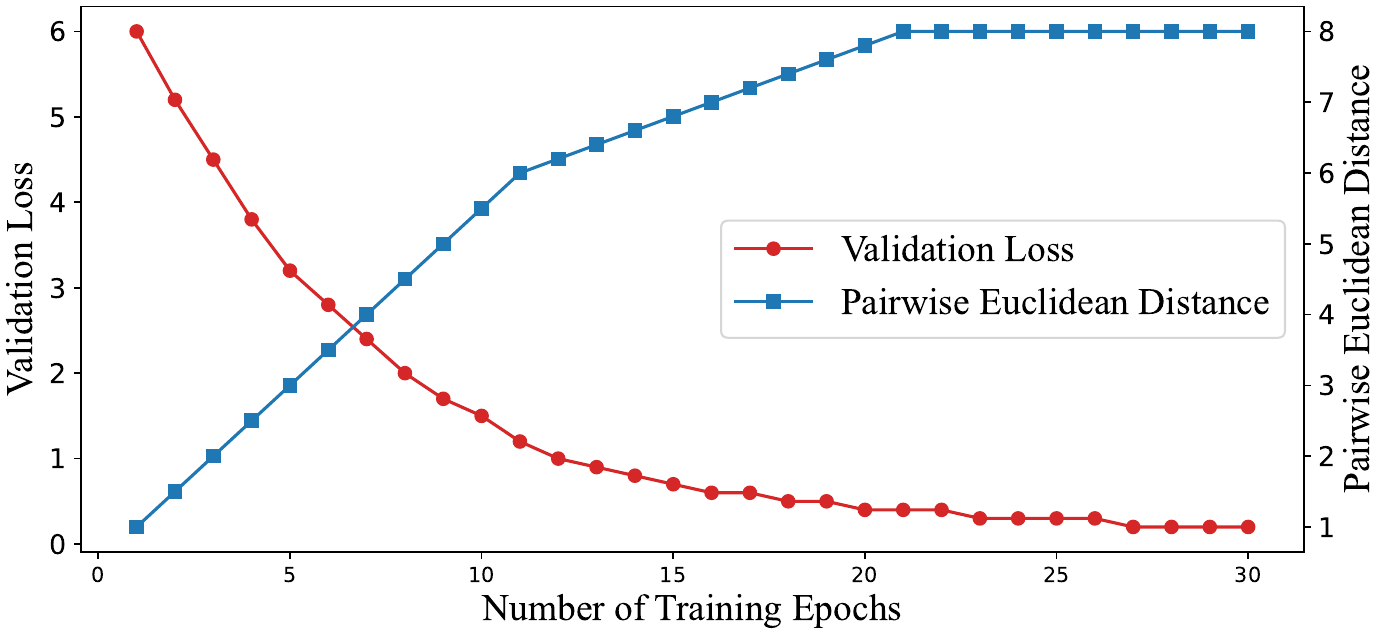}
\vspace{-5pt}
\caption{Validation loss of our channel encoder on the HAT corpus, alongside the average pairwise Euclidean distance between channel embeddings, with respect to the number of training epochs.}
\label{fig:dist}
\vspace{-5pt}
\end{figure}

\subsection{Analysis of Validation Loss and Embedding Distance}\label{sec:dist}

To quantitatively assess the channel discrimination capability of our encoder, we analyzed the evolution of both the validation loss and the average pairwise Euclidean distance between channel embeddings during training. The pairwise distance, computed using the validation set, was averaged between embeddings of the same utterance set (same speaker and content) recorded across different channels. As shown in Fig. \ref{fig:dist}, the validation loss (red curve) decreases steadily over 30 epochs, while the pairwise distance (blue curve) increases, indicating the encoder is effectively learning distinct channel characteristics. This positive trend supports the critical role of the channel encoder in the success of CADA-GAN.

\section{Conclusion and Future Work}\label{sec:conclusion}

This study\footnote{Code: \url{https://github.com/JethroWangSir/CADA-GAN/}.} presents CADA-GAN, a novel method for channel compensation in robust ASR. By integrating a channel encoder with a GAN architecture, CADA-GAN effectively addresses channel mismatch, improving ASR generalization to unseen conditions. Experiments on the HAT and TAT corpora demonstrate that CADA-GAN significantly outperforms strong baselines, achieving substantial CER reductions and higher MOS scores. These results underscore the efficacy of our method in improving both the accuracy and perceptual quality of speech recognition across diverse channel environments.

Future work will involve further validating the effectiveness of CADA-GAN with more advanced ASR models like Whisper\textsubscript{Large} and extending the evaluation to a wider range of challenging datasets. Additionally, we aim to explore integrating CADA-GAN with other domain adaptation techniques to address multiple sources of variability in speech data.

\bibliographystyle{IEEEtran}
\bibliography{references.bib}

\end{document}